\documentclass[a4paper,11pt]{article}
\usepackage{pos}
\usepackage{siunitx}
\usepackage{hyperref}
\usepackage{setspace}

\title{The very high energy view of gamma-ray bursts with the MAGIC telescopes}

\tiny

\author*[a]{Alessio Berti}
\author[b]{Željka Bošnjak}
\author[c]{Alberto Castro-Tirado}
\author[d]{Stefano Covino}
\author[e,f]{Susumu Inoue}
\author[g]{Francesco Longo}
\author[h]{Serena Loporchio}
\author[i]{Davide Miceli}
\author[a]{Razmik Mirzoyan}
\author[]{Elena Moretti}
\author[d]{Lara Nava}
\author[f]{Koji Noda}
\author[a]{David Paneque}
\author[j]{Antonio Stamerra}
\author[k]{Yusuke Suda}
\author[l]{Kenta Terauchi}
\author[e]{Ievgen Vovk}
\author[]{on behalf of the MAGIC Collaboration}
\author[e]{Katsuaki Asano}
\author[m]{Satoshi Fukami}
\author[n]{Nuria Jordana-Mitjans}
\author[d]{Andrea Melandri}
\author[n,o]{Carole Mundell}
\author[j]{Michele Palatiello}
\author[p,q]{Manisha Shrestha}
\author[q]{Iain Steele}

\affiliation[a]{Max Planck Institute for Physics, Boltzmannstrasse 8, 85748 Garching bei München, Germany}
\affiliation[b]{Faculty of Electrical Engineering and Computing, University of Zagreb, Zagreb, Croatia}
\affiliation[c]{Instituto de Astrofísica de Andalucía, Glorieta de la Astronomía s/n, 18008 Granada, Spain}
\affiliation[d]{INAF-Osservatorio Astronomico di Brera, Via E. Bianchi 46, I-23807 Merate (LC), Italy}
\affiliation[e]{Institute for Cosmic Ray Research, The University of Tokyo, Kashiwanoha 5-1-5, Kashiwa, Japan}
\affiliation[f]{Chiba University, ICEHAP, 263-8522, Chiba, Japan}
\affiliation[g]{INFN, Sezione di Trieste, via Valerio 2, Trieste, Italy}
\affiliation[h]{INFN MAGIC Group: INFN Sezione di Bari and Dipartimento Interateneo di Fisica dell’Università e del Politecnico di Bari, I-70125, Bari, Italy}
\affiliation[i]{Dipartimento di Fisica e Astronomia dell’Università and Sezione INFN, Padova, Italy}
\affiliation[j]{National Institute for Astrophysics (INAF), I-00136 Rome, Italy}
\affiliation[k]{Physics Program, Graduate School of Advanced Science and Engineering, Hiroshima University, 739-8526 Hiroshima, Japan}
\affiliation[l]{Department of Physics, Kyoto University, 606-8502 Kyoto, Japan}
\affiliation[m]{DESY, Platanenallee 6, 15738 Zeuthen, Germany}
\affiliation[n]{Department of Physics, University of Bath, Claverton Down, Bath, BA2 7AY, UK}
\affiliation[o]{European Space Agency, European Space Astronomy Centre, 28692 Villanueva de la Canñada, Madrid, Spain}
\affiliation[p]{Astrophysics Research Institute, Liverpool John Moores University, Liverpool Science Park IC2, 146 Brownlow Hill L3 5RF, UK}
\affiliation[q]{Steward Observatory, University of Arizona, 933 North Cherry Avenue, Tucson, AZ 85721-0065, USA}

\emailAdd{aberti@mpp.mpg.de}

\abstract{
Gamma-ray bursts (GRBs) are one of the main targets for the observations of the MAGIC telescopes. As a result of the effort in improving the sensitivity of the instrument and the automatic follow-up strategy, MAGIC detected two GRBs in the very-high-energy (VHE, $E>100$ GeV) range, namely GRB 190114C and GRB 201216C. In GRB 190114C ($z=0.42$), the data collected by MAGIC revealed a new emission component at sub-TeV energies in the afterglow of the GRB. The very rich multi-wavelength dataset, spanning 17 orders of magnitude in energy, allowed to perform a detailed modelling of the broadband emission. The multi-wavelength data could be modelled within a one-zone synchrotron-self Compton scenario with internal $\gamma-\gamma$ absorption, where the model parameters are compatible with those found in previous GRB afterglow studies below GeV energies. Similarly, GRB 201216C broadband emission could be explained using the same model, although the amount of simultaneous multi-wavelength data is reduced with respect to GRB 190114C. In particular, GRB 201216C challenged the current MAGIC detection potential, as its redshift was determined to be $z=1.1$, strongly reducing the observed gamma-ray flux but making it the most distant source detected at VHE. These two detections, accompanied by evidence of VHE emission from a few more GRBs, opened up new questions such as the presence of sub-TeV emission in different classes and phases of GRBs. In this contribution we will present the status of the MAGIC GRB follow-up program, with an highlight on its detected GRBs. Moreover we will show the results on the GRBs observed by MAGIC from 2013 to 2019 with no evidence of VHE emission, in particular for those with simultaneous X-ray observations and redshift $z<2$. We will discuss the implications of these results for GRB physics and the challenges and prospects for future GRB observations with MAGIC.
}

\ConferenceLogo{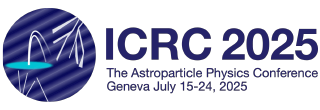}

\FullConference{39th International Cosmic Ray Conference (ICRC2025)\\
 15–24 July 2025\\
Geneva, Switzerland\\}

\begin{document}
\maketitle

\section{Introduction}
\label{sec:intro}

\noindent Gamma-ray bursts (GRBs) are the archetype of transient sources, releasing isotropic equivalent energies up to \SI{e54}{erg} mainly in the \unit{\keV}-\unit{\MeV} range. The bulk of the emission is observed in the so-called \textit{prompt} phase, whose duration can range from fractions of a second to hundreds of seconds. According to their duration, GRBs are commonly classified as long and short, using \SI{2}{\second} as the discrimination value. After the prompt, GRBs also show an \textit{afterglow} phase, a long lasting emission in time that can be detected, for some wavelengths, even up to months after the GRB onset. The prompt and the afterglow are also distinguishable by their time variability properties: the former exhibits flux variability down to milliseconds scale, while the latter is characterized by a smooth decay of the emission with time. Finally, while the prompt is usually detected in the X-ray and MeV energy ranges, the afterglow instead can be detected in a wide range of wavelengths, from radio to GeV gamma rays. \\
Several questions are still open for both the prompt and afterglow phase. Among others, the exact jet launching mechanism and the exact energy dissipation process are prime examples of such open issues. Especially for the prompt, the difficulty in obtaining a more complete picture via multi-wavelength observations due to its intrinsic transient nature is a significant limiting factor. The afterglow instead has been deeply characterized via observations from the radio up to the GeV energy range. In recent years, the detection of very-high-energy emission (VHE, $E\gtrsim\SI{100}{\GeV}$) from the afterglow of a few GRBs has been a major and awaited breakthrough. Four GRBs were detected by imaging atmospheric Cherenkov telescopes (IACTs), GRB~190114C and GRB~201216C by MAGIC \cite{grb190114c_detection,grb190114c_modelling,grb201216c}, and GRB~180720B and GRB~190829A by H.E.S.S \cite{grb180720b,grb190829a}. A fifth GRB, GRB~221009A, has been detected by the LHAASO extensive air shower array \cite{boat_lhaaso}.\\
The discovery of emission at such energies opened a new window in GRB physics, posing the question if such feature is common among GRBs and if a single emission process like the synchrotron self-Compton (SSC) can explain the afterglow phenomenology in the VHE range. In order to give a better insight into the properties of the afterglow radiation at VHE, there is the necessity to pursue more detections of GRBs via dedicated observation programs. In such a context, the MAGIC collaboration has been very active in such searches, with a remarkable statistic of GRB follow-ups and two detections. This is the result of an extensive follow-up program with clear observation strategies selecting the GRBs with the highest potential for detection, as described in the next Section.

\section{The MAGIC GRB follow-up program}
\label{sec:grb_program}

\noindent MAGIC is a system of two identical Cherenkov telescopes of \SI{17}{\meter} diameter located at \SI{2200}{\meter} above sea level in the Roque de Los Muchachos Observatory in La Palma, Canary Islands, Spain. GRBs are one of the main scientific targets of the MAGIC collaboration, pursuing the detection of VHE emission from these sources. The MAGIC telescopes are particularly suited for the observation of GRBs, and for the follow up of transient events in general: they have a low energy threshold ($\sim$ 50 GeV for small zenith angles), a high sensitivity for sub-TeV energies even for short timescales and a fast repointing speed (7 deg/s). These features are crucial considering the expected rapid decay of the emission with time, and the expected absorption of VHE gamma rays due to the interaction with the \textit{extragalactic background light} (EBL).

\begin{figure}[!htb]
	\centering
	\includegraphics[width=0.9\textwidth]{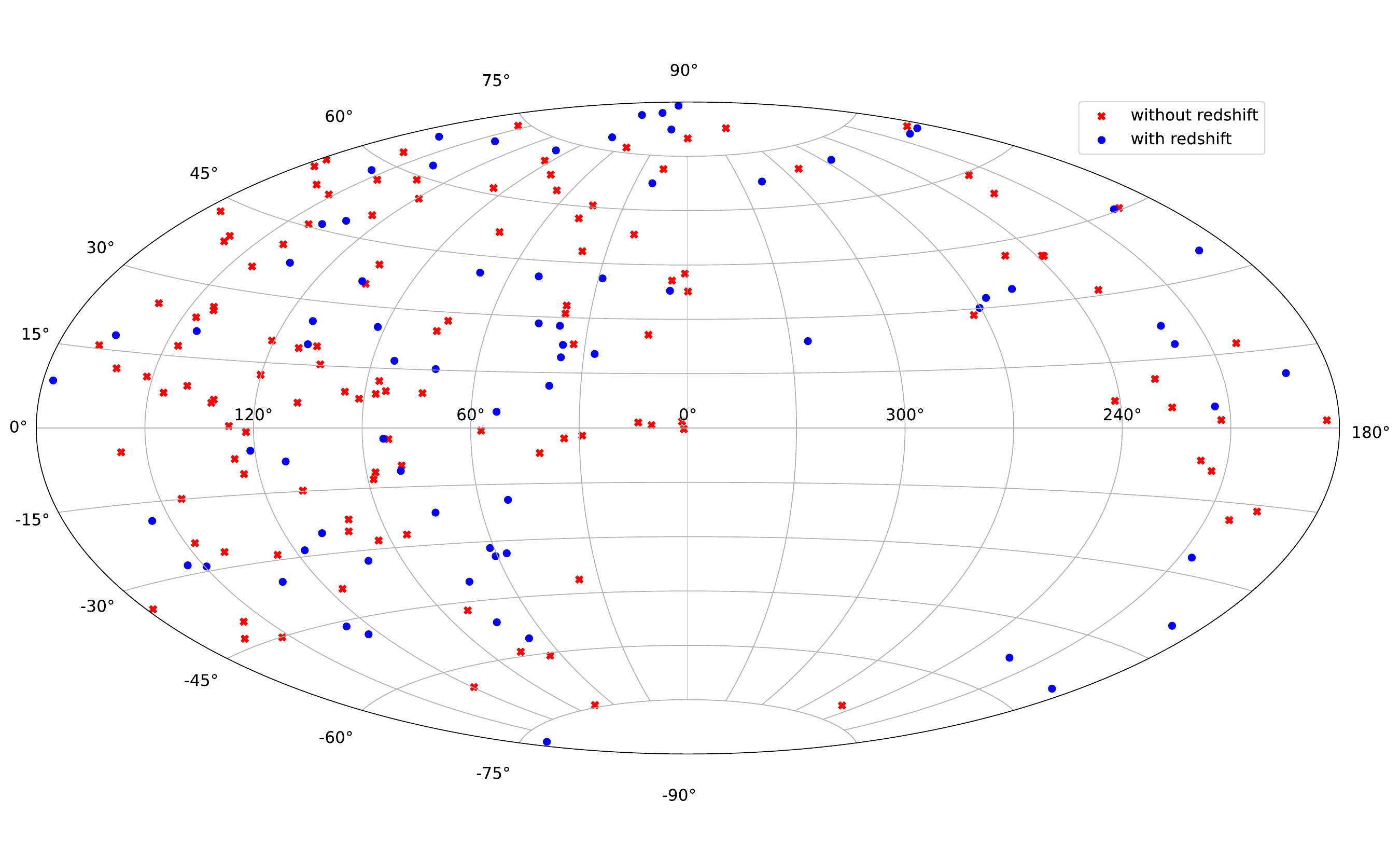}
	\caption{Aitoff projection skymap showing the position of the GRBs followed-up by MAGIC until June 2025. Red crosses denote GRBs without a redshift estimation, while blue dots are GRBs with known redshift.}
	\label{fig:grb_magic_skymap}
\end{figure}

\noindent To ensure a rapid follow-up of GRBs and reduce latency to the minimum, the response to GRB alerts delivered through brokers like the Gamma-ray Coordinates Network (GCN; \url{https://gcn.gsfc.nasa.gov/}) has been fully automatized through the MAGIC automatic alert system (AAS, see \cite{grb190114c_detection}). Alerts received through the AAS are filtered according to their properties (e.g.\ significance, flux) and visibility considerations (e.g.\ maximum zenith, angular distance from the moon etc.). Whenever an alert satisfies the selection criteria, it triggers the automatic reaction of the telescopes, stopping the ongoing observation. The automatic procedure was upgraded in 2013, in particular the data acquisition system is not stopped during the slewing of the telescopes, with a rate limiter avoiding its saturation. This has allowed MAGIC to increase the robustness of the automatic procedure, reducing drastically the number of technical issues during the earliest phases of the follow-up. After that, the follow-up is carried for up to 4h after the beginning of the visibility window, but the observation time can be shortened or increased by the Burst Advocate based on available information about the GRB, e.g.\ redshift, retractions or multi-wavelength follow-ups. \\
While the majority of alerts to which MAGIC reacts to come from the \textit{Swift}-BAT instrument, a smaller fraction comes from \textit{Fermi}-GBM, \textit{Fermi}-LAT and INTEGRAL. Recently, also the SVOM satellite started providing GRB alerts, with a rate similar to the one provided by \textit{Swift}-BAT. The success of the extensive MAGIC GRB follow-up program resulted in the detection of two GRBs, GRB~190114C and GRB~201216C. However, the actual number of GRBs followed-up is much larger, highlighting the challenges that are posed in such searches. Since the beginning of the operations of the MAGIC telescopes, around 200 follow-ups have been performed, of which 74 have known redshift.
It has to be noted that the number of follow-ups increased after 2019. Indeed, at that moment the late-time follow-up that was adopted only for LAT detected GRBs was extended to all GRBs, given the intimate connection between X-ray and GeV-TeV emissions, as demonstrated in the two GRBs detected by MAGIC and described in the next section.

\section{GRB~190114C and GRB~201216C}
\label{sec:detected}

\begin{figure}[!htb]
\centering
\includegraphics[width=0.7\textwidth]{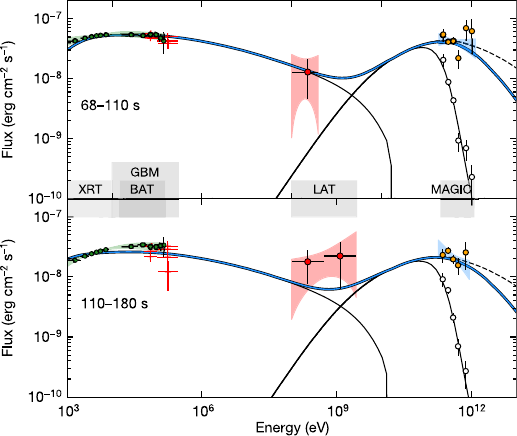}
\caption{GRB~190114C multi-wavelength modeling within the afterglow forward shock scenario, where the MAGIC data is interpreted as SSC emission. Two different time interval are considered. From \cite{grb190114c_modelling}.}
\label{fig:grb190114c_sed}
\end{figure}

\noindent GRB~190114C was the first firm detection of a long GRB by an IACT. The GRB was initially detected by the \textit{Swift}-BAT, and MAGIC started the observations within one minute from the GRB onset. Given the brightness of the GRB, despite the non-optimal observation conditions (moderate moonlight and high zenith), the GRB was detected at a level of $50\sigma$ above \SI{300}{\GeV} in the first 20 minutes of observation. The smoothly decaying behavior of the VHE emission allowed to attribute it to the afterglow phase, and the similar temporal decay as the X-ray emission detected by BAT suggested a connection between the processes responsible for the radiation in those two bands. However, the energies of the gamma rays detected by MAGIC were well above the maximum energy allowed for synchrotron photons emitted by relativistic electrons with the same acceleration and cooling sites. For this reason, GRB~190114C was the first evidence for a new emission component in the afterglow, attributed to the SSC process. The modeling of the multi-wavelength emission can be seen in Fig.~\ref{fig:grb190114c_sed}, where the SSC component at VHE takes into account the internal $\gamma-\gamma$ absortion and the scattering in the Klein-Nishina regime for the intrinsic spectrum, and the absorption of the VHE flux by the EBL for the observed one. While the model was not able to put a preference over a wind-profile or constant density for the circumburst medium, the parameters found are in agreement with the ones found in previous studies of GRB afterglows up to GeV energies and with no VHE detection (see e.g.\ \cite{miceli_nava_review}). This hints to the possibility that the VHE emission in GRBs may be a common feature, and that it can be detected if the observation conditions and the properties of the GRB are favorable.

\noindent A second GRB was detected by MAGIC almost two years later. The long and bright GRB~201216C was a challenging detection due to its distance ($z=1.1$), which translates in a large absorption of the VHE flux due to the EBL. Despite this, MAGIC detected the GRB at the $6\sigma$ level for the first 20 minutes of data above \SI{70}{\GeV}, making GRB~201216C the farthest source detected at VHE. Due to the large EBL absorption, the highest energies detected by MAGIC are around \SI{200}{\GeV}, and this also introduces additional systematic in the spectrum reconstruction, see \cite{grb201216c}. Also in this case, the VHE emission could be modeled within the SSC scenario for the afterglow, with a solution allowed only for the wind density medium. However, given the paucity of multi-wavelength available for this GRB, this translates in a higher degeneracy for the model parameters. Therefore GRB~201216C showcases the capability of IACTs in detecting relatively high redshift GRBs, but it also highlights the importance of having a suitable multi-wavelength dataset allowing a proper interpretation of the emission.

\begin{figure}[!htb]
	\centering
	\includegraphics[width=0.7\textwidth]{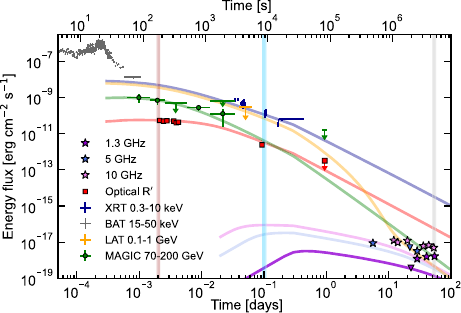}
	\caption{Multi-wavelength light curves of GRB~201216C, showing the radio, optical, X-ray, LAT and MAGIC data. The best fit model in the synchrotron - SSC forward shock scenario is shown as solid lines. See \cite{grb201216c} for details on the model parameters. From \cite{grb201216c}.}
	\label{fig:grb201216c_lc}
\end{figure}

\section{Study of non-detected GRBs}
\label{sec:nondetect}

\begin{figure}[!htb]
\centering
\includegraphics[width=0.8\textwidth]{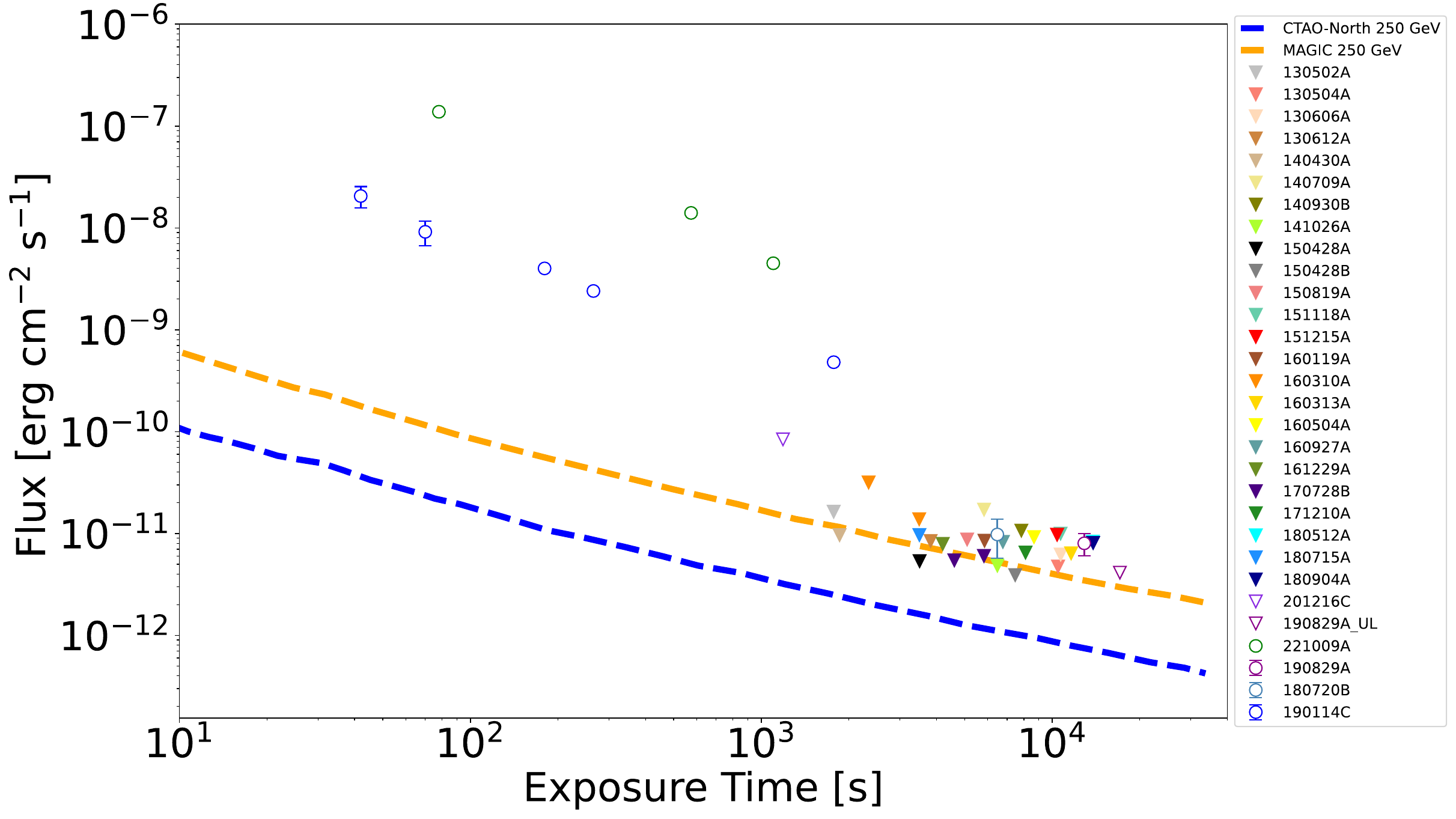}
\caption{Comparison of the MAGIC and CTAO-North array $2\sigma$ sensitivity at 250 GeV (orange and blue dashed lines, respectively) with the observed flux points or ULs for TeV-detected GRBs (empty markers) and the most stringent ULs for the non-detected GRBs by MAGIC (filled markers), as a function of exposure time. From \cite{grb_ul_paper}.}
\label{fig:MAGIC_non_interesting_GRBs_250}
\end{figure}

\noindent While GRB~190114C and GRB~201216C proved that GRBs emit at VHE in the afterglow phase, and that SSC is flexible enough for the interpretation of their emission, additional information can be extracted from the study of non-detected GRBs. These are the vast majority in the sample of GRBs followed-up by MAGIC. A sample of 39 GRBs observed between 2013 and 2019 was selected for this purpose, extracting integral and differential energy flux upper limits (ULs) at 95\% confidence level \cite{grb_ul_paper}. \\
Such sample is quite diverse in terms of observational conditions, in particular observation delay and zenith (the latter affecting the attainable energy threshold) and properties of the GRBs, e.g.\ the availability of redshift measurement and X-ray flux. For this reason, for GRBs with unknown redshift or with redshift $z \geq 2$ or observed at zenith angle Zd $ > 40$ deg (33 out of 39), the nightly flux ULs were compared to the $2\sigma$ level sensitivities of the MAGIC and CTAO-North array at two reference energy values (150 GeV and 250 GeV, respectively), and the fluxes obtained for the GRBs detected in the VHE domain (see Fig.~\ref{fig:MAGIC_non_interesting_GRBs_250} for the comparison at 250 GeV). As a result of this comparison, the GRBs from this sub-sample have intrinsic properties not dissimilar to the ones of the detected GRBs, and the ULs are below the flux values for detected GRBs, lying above the CTAO-North $2\sigma$ sensitivity. Therefore, the undetected GRBs are fainter or at larger redshift, making the detection more difficult, or even not possible.

\begin{figure}[!htb]
\centering
\includegraphics[width=0.7\textwidth]{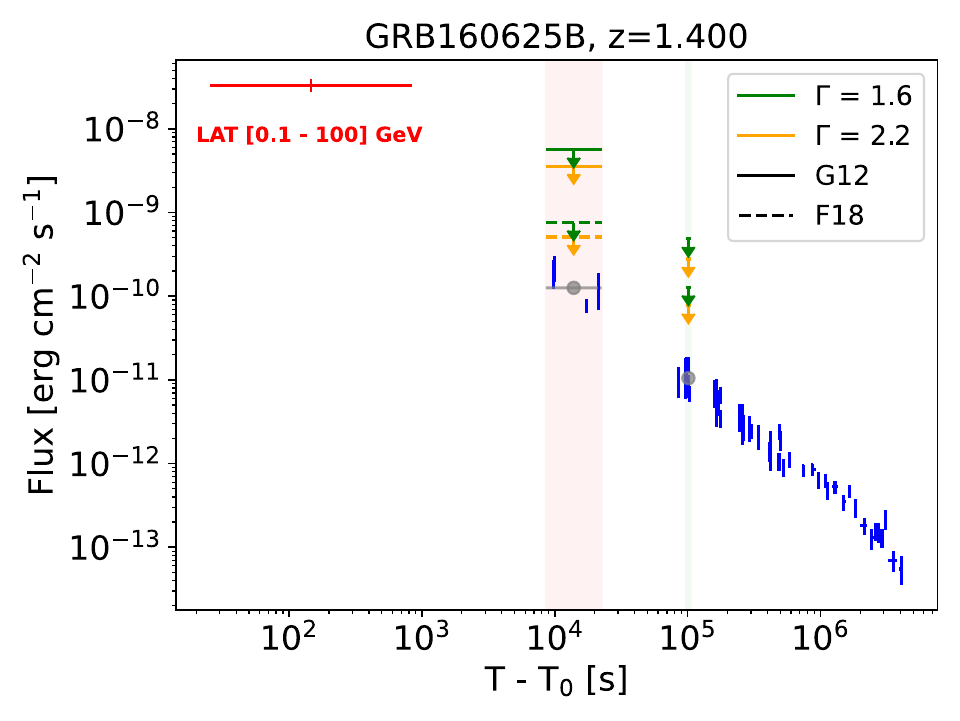}
\caption{Multi-wavelength lightcurve fro GRB~160625B, showing the X-ray fluxes from \textit{Swift}-XRT and its average in the MAGIC observational time windows as grey points, the LAT data in red and the MAGIC ULs. The latter are computed assuming two different photon indices and EBL models. The red and green vertical strips denote the MAGIC observational time windows. From \cite{grb_ul_paper}.}
\label{fig:grb160625b}
\end{figure}

\noindent For the remaining 6 GRBs with redshift $z < 2$ and observed at zenith angle Zd $< 40$ deg, a different analysis was performed by computing the EBL-corrected (de-absorbed) flux ULs in selected energy and time intervals. Moreover, different assumptions for the intrinsic gamma-ray spectrum and EBL absorption models were adopted. The de-absorbed VHE flux was compared with the \textit{Swift}-XRT flux in the soft X-ray band in order to test the connection between this band and the VHE one, and to check for the universality of the VHE emission. These results are shown in Fig.~\ref{fig:grb160625b} for the case of GRB~160625B, where also the \textit{Fermi}-LAT estimated flux in the 0.1 - 100 GeV range is shown. It is found that VHE ULs cannot constrain the TeV component at the same level or below the X-ray one. \\
Therefore, the presence of VHE emission in the sample of undetected GRBs cannot be excluded when considering the MAGIC flux ULs, and its brightness relation to the X-ray emission cannot be firmly established. By comparing the ULs with the sensitivity of CTAO, it is expected to have more GRBs detected at VHE, possibly fainter ones with respect to the ones detected until now, giving the possibility to obtain important information on a large population of GRBs.

\section{Conclusions}
\label{sec:conclusions}

\noindent The MAGIC GRB follow-up program, in place since the beginning of the experiment, lead to a relevant amount of GRB observed in the last 22 years of operation. The continuous efforts in improving and fine-tuning the follow-up strategies culminated in the detection of VHE emission from GRB~190114C and GRB~201216C. For those GRBs, the SSC model was shown to be a plausible interpretative scenario for the sub-TeV emission, leading to the question if such component may be generally adopted for all GRBs detected in the same energy range. Moreover, they hinted to a connection between the luminosities in the VHE and X-ray ranges. Such connection was tested using the data of 39 GRBs not detected by MAGIC between 2013 and 2019, concluding that it cannot be firmly established with the available sample. A larger number of GRBs detected in the VHE range is therefore necessary to gather more information on the properties of the VHE emission. While current generation IACTs like MAGIC can still contribute to the task, the future CTAO with its improved sensitivity is expected to provide more GRB detections, and possibly start population studies for GRBs in the VHE range.

%

\bibliographystyle{JHEP}
\bibliography{grbs_magic_icrc2025}

\section*{Acknowledgements}
\begin{spacing}{1.0}
\noindent {\tiny We would like to thank the Instituto de Astrof\'{\i}sica de Canarias for the excellent working conditions at the Observatorio del Roque de los Muchachos in La Palma. The financial support of the German BMBF, MPG and HGF; the Italian INFN and INAF; the Swiss National Fund SNF; the grants PID2019-107988GB-C22, PID2022-136828NB-C41, PID2022-137810NB-C22, PID2022-138172NB-C41, PID2022-138172NB-C42, PID2022-138172NB-C43, PID2022-139117NB-C41, PID2022-139117NB-C42, PID2022-139117NB-C43, PID2022-139117NB-C44, CNS2023-144504 funded by the Spanish MCIN/AEI/ 10.13039/501100011033 and "ERDF A way of making Europe; the Indian Department of Atomic Energy; the Japanese ICRR, the University of Tokyo, JSPS, and MEXT; the Bulgarian Ministry of Education and Science, National RI Roadmap Project DO1-400/18.12.2020 and the Academy of Finland grant nr. 320045 is gratefully acknowledged. This work was also been supported by Centros de Excelencia ``Severo Ochoa'' y Unidades ``Mar\'{\i}a de Maeztu'' program of the Spanish MCIN/AEI/ 10.13039/501100011033 (CEX2019-000920-S, CEX2019-000918-M, CEX2021-001131-S) and by the CERCA institution and grants 2021SGR00426 and 2021SGR00773 of the Generalitat de Catalunya; by the Croatian Science Foundation (HrZZ) Project IP-2022-10-4595 and the University of Rijeka Project uniri-prirod-18-48; by the Deutsche Forschungsgemeinschaft (SFB1491) and by the Lamarr-Institute for Machine Learning and Artificial Intelligence; by the Polish Ministry Of Education and Science grant No. 2021/WK/08; and by the Brazilian MCTIC, CNPq and FAPERJ.}
\end{spacing}

\end{document}